\documentclass[aps,prd,preprint,groupedaddress,longbibliography]{revtex4-1}

\usepackage{amsmath,amssymb,amsfonts}
\usepackage[T1]{fontenc}
\usepackage{graphicx}
\usepackage{hyperref}

\begin{document}

\title{Neutron-mirror neutron oscillations in stars}


\author{Wanpeng Tan}
\email[]{wtan@nd.edu}
\affiliation{Department of Physics, Institute for Structure and Nuclear Astrophysics (ISNAP), and Joint Institute for Nuclear Astrophysics - Center for the Evolution of Elements (JINA-CEE), University of Notre Dame, Notre Dame, Indiana 46556, USA}

\date{\today}

\begin{abstract}
Based on a newly proposed mirror-matter model of neutron-mirror neutron ($n-n'$) oscillations [Phys. Lett. B 797, 134921 (2019)], evolution and nucleosynthesis in single stars under a new theory is presented. In the new model, $n-n'$ oscillations are caused by a very small mass difference between particles of the two sectors. The new theory with the new $n-n'$ model can demonstrate the evolution in a much more convincing way than the conventional belief. In particular, many observations in stars show strong support for the new theory and the new $n-n'$ model. For example, progenitor mass limits and structures for white dwarfs and neutron stars, two different types of core collapse supernovae (II-P and II-L), synthesis of heavy elements, pulsating phenomena in stars, etc, can all be easily and naturally explained under the new theory.
\end{abstract}


\maketitle

\section{Introduction}

After the big bang nucleosynthesis (BBN) \cite{alpher1948,pitrou2018}, only light elements are formed with about one quarter of $^{4}$He, three quarters of $^{1}$H, and some trace amounts of $^{2}$H, $^{3}$He, and $^{7}$Li due to the missing links of stable nuclei at mass $A =$ 5 and 8. As it turns out, these primordial elements would serve as fuel to form other isotopes in stars when the conditions of high temperature and density can be met. In stars, hydrogen can be further processed into helium via the so-called pp-chain and CNO reactions \cite{bethe1938,bethe1939}. To overcome the mass gaps at $A=$ 5 and 8, however, the triple-alpha reaction via the Hoyle state ($0^+$ at 7.654 MeV in $^{12}$C) \cite{hoyle1954} is needed to start forming $^{12}$C and subsequently other heavier elements.

Such an elegant picture of nucleosynthesis up to carbon has been firmly established while the current understanding of the formation of the heavier elements beyond carbon in stars is not satisfactory and will be challenged in this work. The conventional view of burning between carbon and iron \cite{hoyle1954} is through alpha capture reactions like $^{12}$C($\alpha,\gamma$)$^{16}$O and fusion reactions starting with $^{12}$C+$^{12}$C. Since iron group nuclei are the most bound ones, isotopes beyond iron have to be generated via the slow and rapid neutron capture processes (s-process and r-process) \cite{burbidge1957} under different conditions in stars. Although the studies on neutron capture processes on the heavy nuclei have gained much attention especially after the detection of a neutron star merger event by LIGO and VIRGO \cite{ligoscientificcollaborationandvirgocollaboration2017}, better understanding of the path and nucleosynthesis of the intermediate nuclei and the seed nuclei for s- and r- processes and these processes themselves is still in need.

It is puzzling if we consider that both $^{12}$C($\alpha,\gamma$)$^{16}$O \cite{deboer2017} and $^{12}$C+$^{12}$C \cite{tan2020c} fusion reactions have been measured with much smaller cross sections than desired and the third most abundant isotope in the Universe is $^{16}$O instead of $^{12}$C. In terrestrial planets including Earth, $^{12}$C is surprisingly much rarer compared to abundant or even dominant $^{16}$O. Also intriguingly, studies have shown that s-process has two (main and weak) components \cite{kappeler2011}, r-process nuclei are related to ``high-'' and ``low-frequency'' events \cite{qian2003}, and core-collapse supernovae can be divided into two categories in terms of light curves \cite{smartt2009,faran2014}. Other enigmatic phenomena include progenitor sizes for white dwarfs and neutron stars, carbon-enhanced metal-poor stars (CEMP) in the early Universe \cite{beers2005,carollo2014}, and dramatic oscillatory behavior in stars beyond main sequence such as pulsating variables. All these puzzles in stars indicate possible new physics related to neutrons and have motivated recent development of a new mirror-matter model with neutron-mirror neutron ($n-n'$) oscillations \cite{tan2019}.

Neutron dark decays \cite{fornal2018} or some type of $n-n'$ oscillations \cite{berezhiani2006,berezhiani2009,berezhiani2018a,berezhiani2019a,tan2019} have become a focus of many research efforts recently, at least partly owing to the 1\% neutron lifetime discrepancy between two different experimental techniques \cite{yue2013,pattie2018}. However, the dark decay idea was dismissed shortly by other experimental work 
\cite{tang2018,ucnacollaboration2018} making $n-n'$ oscillations the only possible option. One is referred to Ref. \cite{tan2019} for more detailed discussions on this aspect. In particular, an interesting study of $n-n'$ oscillations in neutron stars \cite{mannarelli2018} combined with a detailed analysis of pulsar timings and detection of gravitational waves \cite{goldman2019} seems to set a very tight constraint on the effect of $n-n'$ oscillations which will be addressed in this work.

Most proposals of the $n-n'$ type of oscillations tried to introduce some sort of very weak and explicit interaction between particles in ordinary and mirror (dark) sectors \cite{foot2004,berezhiani2004,cui2012}. Such an interaction then results in a small mass splitting of $n-n'$ and hence the oscillations. The issue is that it also inevitably makes the oscillations entangled with magnetic fields in an undesirable way due to the nonzero magnetic moment of neutrons. More and more experiments keep pushing its limit to the extreme \cite{serebrov2009,berezhiani2018a} and effectively disfavor such ideas.

A newly proposed model of $n-n'$ oscillations \cite{tan2019}, contrarily, looks at least more viable. It is based on the mirror matter theory (first proposed in Ref. \cite{kobzarev1966}, further developed later in Refs. \cite{blinnikov1982,blinnikov1983,kolb1985,khlopov1991,foot2004,berezhiani2004,berezhiani2006,okun2007,cui2012}), that is, two sectors of particles have similar yet separate gauge interactions within their own sector but share the same gravitational force. Such a mirror matter theory has appealing theoretical features. The mirror symmetry is particularly intriguing as the Large Hadron Collider has found no evidence of supersymmetry so far and we may not need supersymmetry as conventionally understood, at least not below energies of 10 TeV.

The new mirror-matter model that will be applied in this work can consistently explain various observations in the Universe including the neutron lifetime anomaly and dark-to-baryon matter ratio \cite{tan2019}, puzzling phenomena related to ultrahigh-energy cosmic rays \cite{tan2019b}, baryon asymmetry of the Universe \cite{tan2019c}, unitarity of the CKM matrix \cite{tan2019d}, dark energy and the nature of neutrinos \cite{tan2019e}. Furthermore, various laboratory experiments using current technology have been proposed \cite{tan2019d} to test the new model and measure its few parameters more accurately. The model has also been extended into a set of supersymmetric mirror models under dimensional evolution of spacetime to explain the arrow of time and big bang dynamics \cite{tan2020,tan2020a} and to understand the nature of black holes \cite{tan2020b}.

\section{New mirror-matter model and $n-n'$ oscillations}

In this new mirror matter model \cite{tan2019}, no explicit cross-sector interaction is introduced, unlike other $n-n'$ type models. The critical assumption of this model is that the mirror symmetry is spontaneously broken by the uneven Higgs vacuum in the two sectors, i.e., $\langle\phi\rangle \neq \langle\phi'\rangle$, although very slightly (on a relative breaking scale of $\sim 10^{-15} \text{--} 10^{-14}$) \cite{tan2019}. When fermion particles obtain their mass from the Yukawa coupling, it automatically leads to the mirror mixing for neutral particles, i.e., the basis of mass eigenstates is not the same as that of mirror eigenstates, similar to the case of ordinary neutrino oscillations due to the family or generation mixing. Further details of the model can be found in Ref. \cite{tan2019} and further development in Refs. \cite{tan2019e,tan2020,tan2020a}.

The time evolution of $n-n'$ oscillations in the mirror representation obeys the Schr\"{o}dinger equation,
\begin{equation}\label{eq_sch}
i\frac{\partial}{\partial t}\begin{pmatrix}
\phi_n \\
\phi_{n'}
\end{pmatrix}
= H \begin{pmatrix}
\phi_n \\
\phi_{n'}
\end{pmatrix}
\end{equation}
where natural units ($\hbar=c=1$) are used for simplicity, the Hamiltonian $H$ for oscillations in vacuum can be similarly defined as in the case of normal neutrino flavor oscillations \cite{giunti2007a},
\begin{equation}\label{eq_ham0}
H=H_0=\frac{\Delta_{nn'}}{2} \begin{pmatrix}
-\cos2\theta & \sin2\theta \\
\sin2\theta & \cos2\theta
\end{pmatrix}
\end{equation}
and hence the probability of $n-n'$ oscillations in vacuum is \cite{tan2019},
\begin{equation}\label{eq_prob}
P_{nn'}(t) = \sin^2(2\theta) \sin^2(\frac{1}{2}\Delta_{nn'} t).
\end{equation}
Here $\theta$ is the $n-n'$ mixing angle and $\sin^2(2\theta)$ denotes the mixing strength of about $2\times 10^{-5}$, $t$ is the propagation time that is assumed to be much shorter than the neutron $\beta$-decay lifetime, and $\Delta_{nn'} = m_{n2} - m_{n1}$ is the small mass difference of the two mass eigenstates of about $2\times 10^{-6}$ eV \cite{tan2019} or a possible range of $10^{-6} - 10^{-5}$ eV \cite{tan2019c}. Note that the equation is valid even for relativistic neutrons and in this case $t$ is the proper time in the particle's rest frame.

If neutrons travel in medium such as dense interior of a star, the Mikheyev-Smirnov-Wolfenstein (MSW) matter effect \cite{wolfenstein1978,mikheev1985} may be important, i.e., coherent forward scattering with other nuclei can affect the oscillations by introducing an effective interaction term in Hamiltonian,
\begin{equation}\label{eq_hami}
H_I=\begin{pmatrix}
V_{eff} & 0 \\
0 & 0
\end{pmatrix}.
\end{equation}
and the effective potential due to coherent forward scattering can be obtained as
\begin{equation}\label{eq_veff}
V_{eff} = \frac{2\pi}{m_n}\sum_{i} b_i n_i
\end{equation}
where $m_n$ is the neutron mass, $n_i$ is the number density of nuclei of $i$-th species in the medium, and $b_i$ is the corresponding bound coherent scattering length as tabulated in Ref. \cite{sears1992}. Therefore, the modified Hamiltonian in medium can be written as,
\begin{equation}\label{eq_hamm}
H=H_M=\frac{\Delta_{nn'}}{2} \begin{pmatrix}
-\cos2\theta+V_{eff}/\Delta_{nn'} & \sin2\theta \\
\sin2\theta & \cos2\theta-V_{eff}/\Delta_{nn'}
\end{pmatrix}
\end{equation}
and the corresponding transition probability is
\begin{equation}\label{eq_probm}
P_{M}(t) = \sin^2(2\theta_M) \sin^2(\frac{1}{2}\Delta_{M} t)
\end{equation}
where $\Delta_{M} = C\Delta_{nn'}$, $\sin2\theta_M = \sin2\theta/C$, and the matter effect factor is defined as,
\begin{equation}\label{eq_effc}
C = \sqrt{(\cos2\theta - V_{eff}/\Delta_{nn'})^2 + \sin^2(2\theta)}.
\end{equation}

Other incoherent collisions or interactions in the medium can reset the neutron's oscillating wave function or collapse it into a mirror eigenstate, in other words, during mean free flight time $\tau_f$ the $n-n'$ transition probability is $P_{M}(\tau_f)$. The number of such collisions will be $1/\tau_f$ in a unit time. Therefore, the transition rate of $n-n'$ for in-medium neutrons is,
\begin{equation}\label{eq_ratem}
\lambda_{M} = \frac{1}{\tau_f}\sin^2(2\theta_M) \langle\sin^2(\frac{1}{2}\Delta_{M} \tau_f)\rangle.
\end{equation}

Note that the matter effect factor $C$ cancels in Eqs. (\ref{eq_probm}-\ref{eq_ratem}), i.e., the MSW effect is negligible if the matter density is low enough or the propagation time or reset time is short enough (e.g., when other interactions dominate). Another important feature of the matter effect is that the $n-n'$ oscillations can become resonant as in the case of normal neutrino flavor oscillations \cite{mikheev1985}. The resonance condition is $\cos2\theta = V_{eff}/\Delta_{nn'}$, that is, the effective potential $V_{eff}$ is almost equal to the $n-n'$ mass difference since $\cos2\theta \sim 1$ for $n-n'$ oscillations. The condition obviously depends on the unknown sign of the mass difference as well, which could be determined in laboratory measurements proposed in Ref. \cite{tan2019d}. When it resonates, the effective mixing strength is nearly one compared to the vacuum value of $2\times 10^{-5}$.

Similar medium effects could also be caused by the existence of magnetic fields. Unlike some other mirror matter models that are sensitive to weak magnetic fields \cite{berezhiani2009,berezhiani2018a}, the new model used in this work requires a field of $\sim 10^2$ Tesla to be effective. Typical stars do not produce such strong fields \cite{vidotto2014} and the effect is therefore negligible in this study. See Ref. \cite{tan2019d} for further discussion of such effects under super-strong magnetic fields and possible laboratory studies.

\section{Challenging conventional understanding of evolution of stars}

Now we can apply this model to the evolution and nucleosynthesis of stars. In particular, single stars are discussed for simplicity and assumed to be composed of pure ordinary matter initially as it is typical during the formation of inhomogeneities in the early universe and segregation of
ordinary and mirror matter on the scale of galaxies or stars \cite{blinnikov1982,blinnikov1983,kolb1985,khlopov1991}. We will discuss two cases. One is low mass stars ($ < 8 M_{\odot}$) which will eventually die as a white dwarf. The other is more massive stars (between $8-20 M_{\odot}$) that will undergo supernova (SN) explosion where r-process could occur for making half of all heavy elements \cite{qian2003} and leave a neutron star in the end.

For both cases the star burns hydrogen first via the so-called pp-chains and CNO cycles \cite{bethe1938,bethe1939}. This is the longest burning process and can take up to billions of years depending on its initial mass. Then the ashes of the hydrogen burning, $^4$He nuclei, start forming $^{12}$C via the triple-$\alpha$ process \cite{cook1957} at $T=10^8$ K (9 keV in energy). However, that is where the proposed new nucleosynthesis theory starts to part ways with the conventional wisdom.

\begin{table}
\caption{\label{tab_rates} Reaction rates $N_A \langle\sigma v\rangle$ in unit of cm$^3$/mol/s as function of stellar temperature and reaction Q-values are listed for neutron source reactions and the data are taken from JINA REACLIB database \cite{cyburt2010}. The neutron production efficiency factor $f_n$ is defined as the ratio of the neutron mass to the total mass that goes in the reaction.}
\begin{ruledtabular}
\begin{tabular}{c c c c c c c}
T [$10^8$ K] & $^{13}$C($\alpha,n$) & $^{17}$O($\alpha,n$) & $^{18}$O($\alpha,n$) & $^{22}$Ne($\alpha,n$) & $^{12}$C($^{12}$C,n) & $^{12}$C($^{16}$O,n)\\
\hline
1 & 4.2$\times10^{-14}$ & 9.1$\times10^{-20}$ & $1.3\times10^{-34}$ & 1.3$\times10^{-29}$ & 7.8$\times10^{-135}$ & 4.0$\times10^{-78}$ \\
2 & 3.3$\times10^{-8}$ & 2.9$\times10^{-12}$ & 5.8$\times10^{-17}$ & 1.3$\times10^{-16}$ & 1.1$\times10^{-68}$ & 1.6$\times10^{-51}$ \\
5 & 7.7$\times10^{-2}$ & 2.7$\times10^{-4}$ & 2.4$\times10^{-5}$ & 1.0$\times10^{-6}$ & 3.6$\times10^{-28}$ & 3.8$\times10^{-29}$ \\
10 & 2.5$\times10^{2}$ & 2.0 & 1.3 & 6.3$\times10^{-2}$ & 9.4$\times10^{-14}$ & 1.4$\times10^{-17}$ \\
\hline
Q-value [MeV] & 2.216 & 0.587 & -0.697 & -0.478 &-2.598 & -0.424 \\
\hline
$f_n$ & $\frac{1}{17}$ & $\frac{1}{21}$ & $\frac{1}{22}$ & $\frac{1}{26}$ & $<\frac{1}{24}\times10$\% & $\frac{1}{28}\times10$\% \\
\end{tabular}
\end{ruledtabular}
\end{table}

All the above processes do not produce neutrons. So we first review all the possible nuclear reactions for neutron production in stars. The reaction has to be of $(X,n)$-type where $X$ may be one of existing nuclei like proton, $\alpha$, or $^{12}$C at this moment. It has to be energy-releasing, i.e., with a positive Q-value. Some reactions with a slightly negative Q-value (e.g., $>-1$ MeV) may contribute as well, especially at higher temperatures. Reaction rates of such reactions are taken from JINA REACLIB database \cite{cyburt2010} and listed in Table \ref{tab_rates} where two reactions with positive Q-values immediately stand out,
\begin{eqnarray}\label{eq_nsource1}
^{13}C + \alpha \rightarrow \, ^{16}O + n \\
\label{eq_nsource2}
^{17}O + \alpha \rightarrow \, ^{20}Ne + n
\end{eqnarray}
where the first one is fairly well studied \cite{heil2008} while the second reaction is not, especially at low temperatures \cite{best2013,mohr2017}. As shown in Table \ref{tab_rates}, the neutron production efficiency factor $f_n$ defined as the ratio of the neutron mass to the total mass involved in the reaction will be used extensively in the following discussion.

Conventional understanding for massive stars believes that the density and temperature are high enough at the end of the $3\alpha$ process so that it can start the $^{12}$C + $^{12}$C fusion reaction, subsequently fusing the resulting heavier nuclei like oxygen, silicon, etc, and eventually making the most bound iron material in the core \cite{rolfs1988}. In this scenario, although refuted by the proposed new theory, both $^{12}$C($^{12}$C,n) and $^{12}$C($^{16}$O,n) could play a role in neutron production in stars. Unfortunately, only up to 10\% of their total cross sections (with more than 90\% going to the emission of protons or alphas instead) \cite{bucher2015,fang2017} produce neutrons making the efficiency factor $f_n$ (shown in Table \ref{tab_rates}) too small to contribute. Also listed in Table \ref{tab_rates}, $^{22}$Ne($\alpha,n$) has been considered as the neutron source reaction for the weak s-process in massive stars \cite{kappeler2011}.

Now let us first see how the $n-n'$ oscillation mechanism works in the conventional picture of nucleosynthesis in low mass stars like our sun. According to the conventional understanding, the star may continue to burn some of $^{12}$C to $^{16}$O by alpha capture reaction but it can not start carbon + carbon fusion due to insufficient density and temperature \cite{rolfs1988}. The star now has an envelope and burning shells of H and He mixed with CNO elements and a C/O core and eventually at a stage called asymptotic giant branch (AGB) where s-process occurs for making heavy elements \cite{burbidge1957}. The neutron source reaction
$^{13}$C($\alpha,n$) operates at the outer layer of the star and $^{13}$C can be created from $^{12}$C via $^{12}$C($p,\gamma$)$^{13}$N($\beta^+$)$^{13}$C.

The s-process environment is typically regarded as follows: density $\rho \sim 10^3$ g/cm$^3$; temperature $T \sim 10^8 K$; neutron number density $n_n \sim 10^8$ /cm$^3$ \cite{meyer1994a}. For simplicity, we assume the star has a little iron with a solar abundance that will serve as seed at the start of the s-process.

The mean free flight time $\tau_f$ of neutrons in the stellar medium is determined by the scattering cross sections of nuclei. It can be defined by the scattering rate $\lambda_f$ as follows,
\begin{equation}\label{eq_tauf}
\frac{1}{\tau_f} \equiv \lambda_f = \sum_{\text{all nuclei}} \rho N_A \langle\sigma_{nN}v\rangle Y_N
\end{equation}
where $N_A$ is the Avogadro constant, $\langle\sigma_{nN}v\rangle$ is the thermal average of neutron-nucleus scattering cross section times neutron velocity, and $Y_N$ is the mole fraction of the nucleus (i.e., its mass fraction divided by the mass number of the nucleus) \cite{rolfs1988}.
The typical neutron-nucleus scattering cross section is about one barn as it is dominated by the neutron scattering length for low energy neutrons of $\sim 10$ keV. And the neutron velocity under the s-process temperature ($10^8$ K) is about $1.3\times10^6$ m/s. 

In the outer layer of the AGB where $^{13}$C($\alpha,n$) operates, the sum of $Y_N \sim 0.1$ is typical assuming that most of it is made of helium and CNO elements. Therefore, we can easily get $\tau_f \sim 10^{-9}$ s from Eq. (\ref{eq_tauf}) for neutrons in the s-process environment and the propagation factor of Eq. (\ref{eq_ratem}) is averaged to 1/2 if we omit the matter effect for now. 

On the other hand, we also need to calculate the neutron loss rate due to the capture reactions on heavy nuclei which was the main motivation in the study of the s-process. Similar to Eq. (\ref{eq_tauf}), we can write the neutron loss rate from capture reactions as follows,
\begin{equation}\label{eq_lamcap}
\lambda_{cap} = \rho N_A \langle\sigma_{cap}v\rangle Y_N
\end{equation}
where the neutron capture reaction rate $N_A \langle\sigma_{cap}v\rangle$ is about $10^3$ cm$^3$/mol/s for $^{12}$C and about $10^6$ cm$^3$/mol/s for $^{56}$Fe at s-process temperature \cite{cyburt2010}. For capture reaction on $^{56}$Fe which represents the seed for s-process with $Y_{56Fe} \sim 10^{-5}$ inferred from the solar abundance, the rate $\lambda_{cap}(^{56}Fe)$ is about $10^4$ s$^{-1}$. The rate is similar for capture reactions on light C/O nuclei. However, this capture process does not contribute to the loss rate of neutrons since the resulting $^{13}$C will release the neutron via ($\alpha$,n) reaction later. Therefore, the neutron loss rate due to capture reactions or s-process is $\lambda_{cap} \sim 10^{4}$ s$^{-1}$.

From Eqs. (\ref{eq_ratem}) and (\ref{eq_lamcap}), we can obtain the branching ratio of the neutrons that oscillate into mirror neutrons to those that are captured into nuclei on the condition that the matter or medium effect in Eqs. (\ref{eq_hami}-\ref{eq_effc}) is omitted,
\begin{equation}
Br(\frac{nn'}{cap}) = \frac{\lambda_f}{2\lambda_{cap}} \sin^2(2\theta) \sim 1
\end{equation}
which indicates that similar amounts of neutrons lost to either $n-n'$ oscillations or s-process in the beginning. Note that this branching ratio does not depend on the density because the individual rates depend on the density in the same way and get canceled for the ratio. Also note that the condition is for the very beginning of s-process. The s-process is a very slow process as it has to wait for many long-lived nuclei to decay along the path before it can capture neutrons again \cite{rolfs1988}. So on average, s-process may only use a small fraction of all available neutrons and most of the neutrons may go via the $n-n'$ oscillation process. Additionally, current model simulations \cite{lugaro2001} typically use very small amounts of $^{13}$C ($10^{-6} - 10^{-5} M_{\odot}$) to reproduce the s-process. This shows evidence that $n-n'$ oscillations may take away most of produced neutrons.

Now we can re-visit the oscillation rate considering the matter effect for the following conditions: density of $10^3$ g/cm$^3$ with compositions of 10\% hydrogen and 90\% carbon in mass, scattering lengths of b($^1$H) = -3.74 fm and b($^{12}$C) = 6.65 fm \cite{sears1992}. Then the effective potential can be calculated as $V_{eff} \sim 2\times 10^{-5}$ eV. If we assume that the 90\% part is made of both carbon and oxygen evenly, we can obtain $V_{eff} \sim 6\times 10^{-6}$ eV that is amazingly close to the estimate of the $n-n'$ mass difference of $6.578\times 10^{-6}$ eV assuming equivalence of the
$CP$ violation and mirror symmetry breaking scales \cite{tan2020d}. In fact, in slightly outer regions with lower density of about $10^2$ g/cm$^3$, or for a possible larger $n-n'$ mass splitting up to $10^{-5}$ eV \cite{tan2019c}, $V_{eff}$ and $\Delta_{nn'}$ could be almost identical, i.e., leading to maximal or resonant oscillations. If resonant n-n' oscillations indeed occur, then we can learn that the sign of $\Delta_{nn'}$ is positive.

Then where do the mirror neutrons go? Taking the similar step as suggested in Ref. \cite{mannarelli2018}, the mirror neutrons converted from the oscillations will travel to the core of the star due to gravity. The $n-n'$ oscillations are forbidden in bound nuclei due to energy conservation, but they do occur in stars when neutrons are produced free. However, the neutrons emitted from $^{13}$C($\alpha,n$) can have energy up to 2.2 MeV and potentially escape from the star if it oscillates immediately into a mirror neutron. Fortunately, the very short mean free flight time discussed above makes the neutron thermalized first before oscillating into a mirror neutron. Its thermal energy is about 8.6 keV at $T=10^8$ K. During the thermalization process, the light neutrons (compared to heavy nuclei) could diffuse into outer regions and probably meet the resonant condition and then maximally oscillate into mirror neutrons as discussed above. Assuming that the inner part of the star is white-dwarf-like (e.g., $1 M_{\odot}$ and Earth-size), it can provide a gravitational binding energy of $\sim 0.2$ MeV in addition to the energy the outer layer can supply should the mirror neutron escape. Therefore, most of the mirror neutrons will go to the core.

Note that mirror neutrons interact with ordinary matter only via gravity, so they become uniformly mixed with ordinary matter in the core with equal density. The details on the core evolution will be discussed with the new theory later.

One observation on the factor $f_n$ in Table \ref{tab_rates} seems to be particularly interesting. $^{13}$C($\alpha,n$) converts about 1/17 of the total mass into neutrons. Suppose that all the neutrons oscillate to mirror neutrons ending up in the core, it means that almost 6\% of the star mass will go into the core in this way. Note that other similar reactions contribute as well. This may provide a link to connect the Chandrasekhar limit \cite{chandrasekhar1931} to the mass limit on the progenitor \cite{smartt2009} and will be explored further in the next section.

If this indeed is the scenario, our understanding of stellar nucleosynthesis has to be changed. The CNO elements may have additional functions other than serving as catalyst for making helium. In particular, the CNO elements $^{13}$C and $^{17}$O can trigger $n-n'$ oscillations via ($\alpha,n$) reaction (with positive reaction Q-values). To a certain extent, $^{18}$O($\alpha,n$) and $^{18}$O($\alpha,\gamma$)$^{22}$Ne($\alpha,n$) (with a little negative reaction Q-values) at higher temperatures and other heavier ($\alpha,n$) reactions like $^{21}$Ne($\alpha,n$) (with positive reaction Q-values) at later stages may contribute as well.

\section{New picture of stellar evolution with $n-n'$ oscillations}

As shown below in the proposed new theory, the neutron production process plays a critical role in the evolution and nucleosynthesis of a star. The $n-n'$ oscillations dictate how the degenerate core is formed, how the mass of the progenitor is related to the Chandrasekhar limit and the neutron star mass limit, and possibly why or when the star may explode - a difficult task for current simulations to do.

In the first burning step after the $3\alpha$ process, starting with $^{13}$C($\alpha,n$), $^{16}$O will be accumulated as ashes from the burning of all carbon nuclei. Then in the second step, hydrogen fuel is added and
$^{16}$O($p,\gamma$)$^{17}$F($\beta^+$)$^{17}$O will convert $^{16}$O into $^{17}$O. The second neutron source reaction $^{17}$O($\alpha,n$) starts to take effect and converts all oxygen nuclei to neon nuclei. From both reactions, it effectively converts star matter into mirror neutrons by (1/17 + 1/21) = 10\% according to the $f_n$ factors shown in Table \ref{tab_rates}. At the same time, both neutron source reactions could provide a small fraction of neutrons for the s-process. To meet the Chandrasekhar limit of about $1.4 M_{\odot}$ for a white dwarf, mirror neutrons cannot exceed $0.7 M_{\odot}$ in mass or no more than $7 M_{\odot}$ star matter can be burned. There is another $0.7 M_{\odot}$ of ordinary matter in the core that does not participate in the burning. This sets the higher mass limit of $7.7 M_{\odot}$ for the progenitor of a white dwarf, or the lower mass limit for the progenitor of a core-collapse supernova, which is in excellent agreement with the observation limit of $8\pm1 M_{\odot}$ \cite{smartt2009}.

\begin{figure}
\includegraphics[scale=0.45]{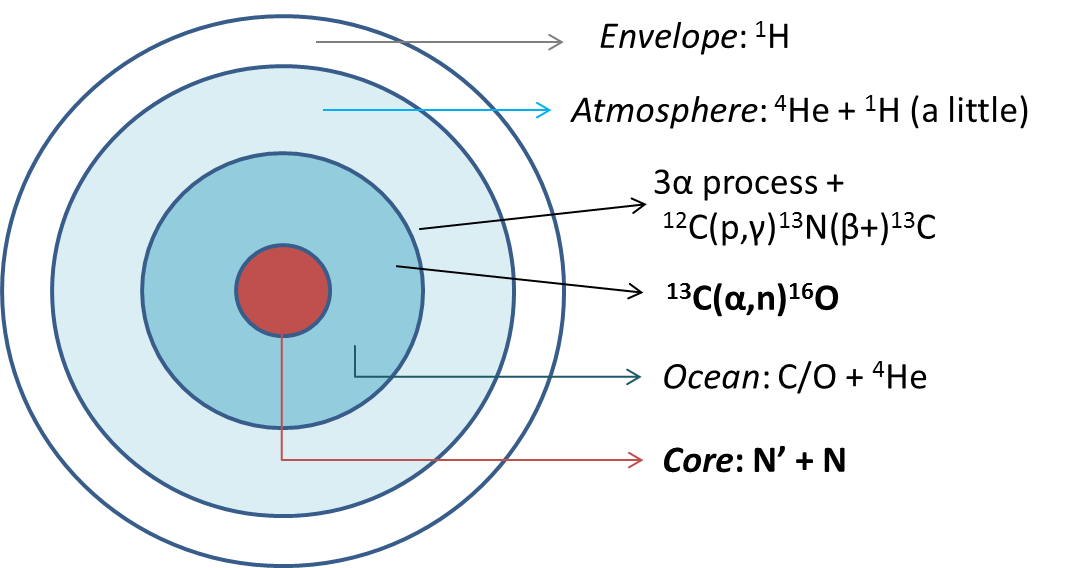}
\caption{\label{fig_1} The schematic diagram is shown for the structure of a red giant star at the first neutron-production $^{13}$C($\alpha,n$) phase. $N$ and $N'$ in the core stand for evenly mixed matter and mirror matter, respectively.}
\end{figure}

As a matter of fact, the above picture is not unlikely and it is more natural. Taken into account the rates from Table \ref{tab_rates} at $T=10^8$ K when the triple-$\alpha$ process starts, one can see how this could occur. At this moment the star as a red giant has a helium core and hydrogen envelope and a small amount of hydrogen is mixed in the helium core. The first step considered here is dictated by the slowest triple-$\alpha$ reaction. Since this burning process is ignited at the center of the core and gradually moved outwards, the red giant becomes brighter as it evolves. The typical structure of the star at this phase is shown in Fig. \ref{fig_1}.

When three helium nuclei fuse into a $^{12}$C nucleus, it quickly captures a mixed-in proton to become unstable $^{13}$N which has a 10-minute $\beta^+$-decay half-life \cite{nndc}. A possible alternative path via $^{12}$C($\alpha,\gamma$) (as commonly believed) does not play a role as its reaction rate is 15 orders of magnitude \cite{cyburt2010} lower than that of $^{12}$C($p,\gamma$) due to a higher Coulomb barrier. Neither does $^{13}$C($\alpha,\gamma$). The only requirement is the existence of a small amount of hydrogen. Several scenarios indeed make it plausible. First, for a low metallicity star, i.e., no significant amount of CNO elements present in its initial composition, only pp-chain burns the initial hydrogen. At this point of the star's life, it is probably no more than or close to two times the p-p reaction lifetime. Therefore, we could have more than 10\% hydrogen left in the core. Even if significant CNO elements exist and exhaust hydrogen in the core, their highly temperature sensitive reaction rates result in plenty of hydrogen left at lower temperature regions outside the core. The core is not degenerate for stars with $M > 2 M_{\odot}$ \cite{pagel1997}, and the burning can cause convection which could bring in fresh hydrogen from the exterior. If none of the above works, when the triple-$\alpha$ burning front grows out of the small original core $^{12}$C($p,\gamma$) and subsequently $^{13}$C($\alpha,n$) can then proceed.

The two reasons why $^{13}$N waits for its decay instead of capturing another proton: most of the nearby hydrogen has been used up first by $^{12}$C; the $^{12}$C($p,\gamma$) rate ($\sim 10^{-5}$ cm$^3$/mol/s) is much higher than that of $^{13}$N($p,\gamma$) ($\sim 10^{-6}$ cm$^3$/mol/s) \cite{cyburt2010}. In the end, $^{13}$N will decay into $^{13}$C. If hydrogen is overabundant in the burning region, the CNO cycles will quickly fuse the excess into $^4$He that are then burned into $^{12}$C and eventually $^{13}$C. Because the $^{13}$C($\alpha,n$) rate is ten orders of magnitude higher than the triple-$\alpha$ rate \cite{cyburt2010}, $^{13}$C is quickly converted into $^{16}$O after the $^{13}$N decay on a 10-minute time scale behind the triple-$\alpha$ burning front. Note that the $n-n'$ oscillations effectively make $^{13}$C($\alpha,n$) a cooling reaction by losing the kinetic energy of the mirror neutron, which may help stabilize the burning front.

As discussed earlier, the generated neutrons then oscillate into mirror neutrons that will go in the core mixing evenly with $^{16}$O. In addition, some of the mirror neutrons actually oscillate back to ordinary neutrons according to Eqs. (\ref{eq_probm}-\ref{eq_ratem}). To calculate the oscillation probability, we assume that, at a later stage, the core as progenitor of a white dwarf has a similar density ($10^6$ g/cm$^3$), where mirror neutrons can be regarded as a gas of free moving particles governed solely by gravity. Applying the virial theorem on the $n'$ system, one can estimate the mean velocity of mirror neutrons $v'=2.5\times10^{-3} (M'/[\text{g}])^{1/3}$ cm/s which grows as the mirror matter mass $M'$ increases. At some stage, e.g., $M'=0.1M_{\odot}$, one can obtain $v' = 1.5\times10^8$ cm/s and hence $\tau'_f \sim 10^{-14}$ s and $\lambda_{n'n} \sim 0.1$ s$^{-1}$. At earlier stages, this reverse oscillation rate can be several orders of magnitude faster. What it does is it provides free neutrons to make the ordinary core material more neutron-rich.

Initially $^{16}$O in the core can be enriched up to its dripline nucleus $^{24}$O \cite{tarasov1997} via $n'\rightarrow n$. Note that these highly neutron-rich nuclei can not undergo the usual beta decays owing to electron degeneracy in the core.
As found out recently, light neutron-rich nuclei have much higher fusion cross sections than normal ones \cite{desouza2017}. So these enriched oxygen nuclei likely fuse further into other neutron-rich intermediate nuclei between oxygen and iron or may further capture the leftover helium near the bottom of the ocean as shown in Fig. \ref{fig_1}, at the same time releasing large amounts of energy. Eventually the core may develop into an onion-like structure starting from the outside layer of O, then Ne, Si, S, Cr, up to Fe in the center. When the temperature in the core is high enough as the mass is close to the Chandrasekhar limit at late burning stages, the core, at least the inner part, may reach a state of nuclear statistical equilibrium (NSE) consisting of mostly iron-group elements with a crust of lighter neutron-rich nuclei.

Alternatively, mirror neutrons can undergo mirror $\beta$-decay $n'\rightarrow p'+e^{'-}+\bar{\nu}'$ with the same lifetime of about 888 sec \cite{tan2019}. When the ordinary core matter is fully enriched, i.e., no more neutrons can be taken, mirror neutrons have to decay to mirror protons. A mirror proton will fuse immediately with a mirror neutron to make a mirror deuteron. Subsequently, the mirror core matter will conduct mirror nucleosynthesis similar to the ordinary one, e.g., three mirror alphas fuse into one mirror $^{12}$C. At the same time, the fusion process on the ordinary side will produce free neutrons that can oscillate into mirror neutrons to enrich the mirror matter. Through this mutual oscillation process, both ordinary and mirror matter will develop into similar evenly-mixed core structures (possibly an iron core at NSE with a neutron-rich crust in the end) as shown in Fig. \ref{fig_1}.

The degenerate core's pressure is maintained by both the electron degeneracy and the large energy release of about 8 MeV per neutron due to nuclear binding energy as the free neutron from the $^{13}$C($\alpha,n$) reaction is effectively and ultimately converted into the nuclidic matter in the core. The Chandrasekhar limit for mixed degenerate ordinary and mirror matter is smaller than the usual value by a factor of $\sqrt{2}$, which is consistent with the observed lower mass limit of $\sim 1 M_{\odot}$ for neutron stars \cite{kiziltan2013}. But large amounts of energy release (much larger than $\lesssim 0.5$ MeV per nucleon of what conventionally believed fusion reactions can provide near the core) could make the limit significantly higher and cause the spread of the neutron star mass distribution. Therefore the average limit could still be similar, i.e., close to the observed average neutron star mass of $1.4 M_{\odot}$.

Once all the helium are exhausted or its density is lowered enough to not sustain the triple-$\alpha$ process, therefore no more $^{13}$C($\alpha,n$) running, the core stops growing.  Without the heat from the burning and the neutron-conversion process in the core, the core begins contraction and cools down, pushing away the red giant's hydrogen envelope.

When the core settles and starts pulling back the hydrogen envelope, it may go into the observed AGB phase, i.e., the second burning step that will be discussed below.

\begin{figure}
\includegraphics[scale=0.45]{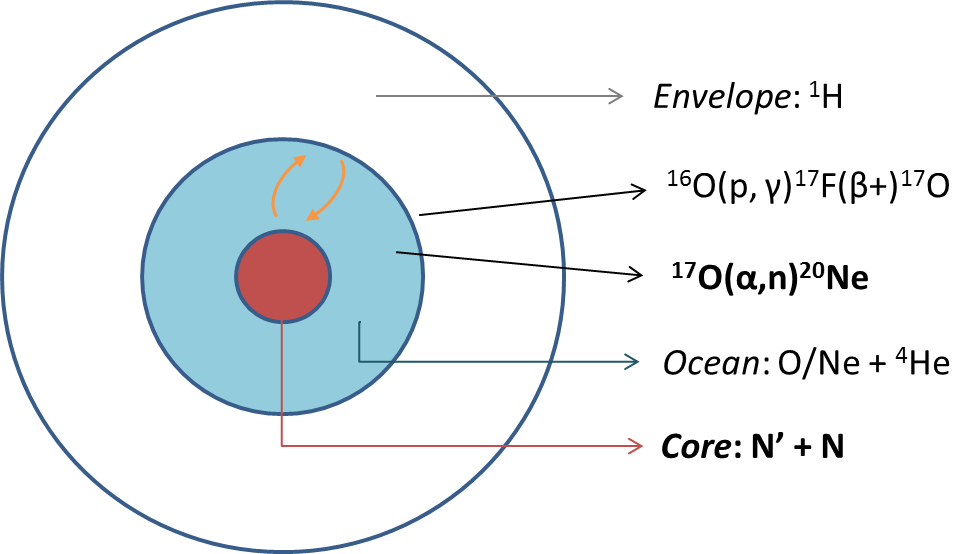}
\caption{\label{fig_2} The schematic diagram is shown for the structure of an AGB star at the second neutron-production $^{17}$O($\alpha,n$) phase. $N$ and $N'$ in the core stand for evenly-mixed matter and mirror matter, respectively.}
\end{figure}

At the second phase, the outer envelope of hydrogen starts falling in and becoming compressed on the surface, it can react with the $^{16}$O on the surface that was newly formed in the previous step and still mixed with some helium. The $^{16}$O($p,\gamma$)$^{17}$F reaction makes $^{17}$F nuclei very quickly, which will sink down in the ocean and decay into $^{17}$O with a 64.5-second $\beta^+$-decay half-life \cite{nndc}. Then the second neutron source reaction of $^{17}$O($\alpha,n$) starts, although at a slower rate than the $^{13}$C($\alpha,n$) rate in the first step. The rate of the only possible competing reaction $^{17}$O($\alpha,\gamma$) is 16 orders of magnitude lower at $T=10^{8}$ K as shown by the work of Best \textit{et al.} \cite{best2013}. The typical structure of the star at the second or AGB phase is shown in Fig. \ref{fig_2}.

Note that the difference here is that the second phase burning starts from just outside and without the helium atmosphere. This probably explains why the AGB stars appear very bright. There may be convection in the ocean to move heavy ash nuclei $^{20}$Ne down and bring $^{16}$O back up. However, it is not required since the heavy $^{20}$Ne sinks into the core, exposing the $^{16}$O to the envelope again as if the envelope ``eating'' away the ocean layer by layer.  Eventually the ocean material outside the core will be all processed in this way. Once no more neutrons are produced, the heat from the neutron-conversion process of $n'-n$ oscillations in the core stops. The star begins contraction again and becomes a white dwarf composed of evenly mixed ordinary-mirror matter in the end.

If the produced mirror neutron matter exceeds $0.7 M_{\odot}$ during the above two steps, in other words, the core weighs beyond the Chandrasekhar limit of about $1.4 M_{\odot}$, the red giant will undergo supernova explosion.
As discussed above, the star will need at least $7.7 M_{\odot}$ as a progenitor to explode. Before the explosion, the $^{13}$C($\alpha,n$) and $^{17}$O($\alpha,n$) reactions in the two phases naturally provide the neutron sources for the main (slower but longer) and weak (faster but shorter) s-processes, respectively. After the explosion, the neutron-rich crust material could be ejected and provide high neutron flux for r-process, which could explain the abundances of r-process nuclei in early generation of stars and diverse sources for r-process \cite{qian2003} as discussed below.

As for the fate of more massive stars with $M > 8M_{\odot}$, there may actually be double core collapses for ordinary and mirror matter, respectively. The ordinary and mirror matter will become a mixture of mirror and ordinary neutrons forming the $n-n'$ star. As shown above, the core of the star can exceed the Chandrasekhar limit during any of the two phases. So we should see two types of core-collapse supernovae that can actually be identified with the observed ones. The cores formed in both cases are essentially the same while the outer layers are much different and can help distinguish the two types.

First, Type II-Plateau supernovae (SNe II-P) have been reported with the following properties \cite{smartt2009,faran2014}: most common (60\%); less peak brightness but with a plateau in light curve; progenitor of $8-15 M_{\odot}$; strong hydrogen lines with no helium. This matches exactly the type of supernovae collapsed in the second phase. Considering $f_n = 1/17$ for the first step reaction $^{13}$C($\alpha,n$) as shown in Table \ref{tab_rates}, the star needs to burn at most $12 M_{\odot}$ to go through the first step without reaching the Chandrasekhar limit. Adding $1 M_{\odot}$ in the unburned ordinary core and $2 M_{\odot}$ for the outer layers, one gets the upper mass limit of $15 M_{\odot}$. Combined with the lower mass limit from the white dwarf analysis above, this type indeed matches the same mass range for the less massive supernovae. During the second step, the burning starts from outside making the ocean layer very thick. When the core collapses, it has to blow off the thick O/Ne ocean layer which will lower its peak luminosity. On the other hand, during the explosion, the thick O/Ne layer may continue to generate energy by nucleosynthesis and therefore present itself as the plateau in light curve. The helium atmosphere is gone after the first step, and the hydrogen envelope is participating directly in burning during the second step, explaining why hydrogen spectrum lines are strong but no evidence of helium. This type of SNe may be the ``high-frequency'' events for heavy r-process nuclei \cite{qian2003}.

Second, Type II-Linear supernovae's (SNe II-L) features are as follows \cite{faran2014}: relatively rare (a few percent); more peak brightness but linear decline in light curve; progenitor more massive ($>15 M_{\odot}$); evidence of helium; hydrogen lines appearing later and weaker. This matches exactly the type exploded in the first phase. The very slow triple-$\alpha$ reaction starts the burning from the core. The subsequent neutron production reaction is much faster, growing the core accordingly. Therefore, the ocean layer is very thin. When the star explodes, it just needs to blast away the light helium atmosphere. The result is more luminosity in the peak and also a quick decline in light curve. Explosions in the first phase need more mass as discussed above. During the triple-$\alpha$ burning, the hydrogen envelope was pushed away and hence producing weaker hydrogen lines at a later time. This type of SNe may be the ``low-frequency'' events for light r-process nuclei \cite{qian2003}. This type of more massive SNe may also dominate in the early universe as they evolve faster and large amounts of neutrons ejected during the explosion can quickly burn the helium layer into carbon via $^4$He+$^4$He$+n \rightarrow ^9$Be and $^9$Be($\alpha,n$)$^{12}$C reactions that are much faster than the triple-$\alpha$ process \cite{delano1971}. This may enhance the carbon abundance in the early generation of stars leading to the so-called carbon-enhanced metal-poor (CEMP) stars \cite{carollo2014}.

Neutron star progenitors with mass beyond $20 M_{\odot}$ are rarely observed \cite{smartt2009}. Under this theory, we may be able to obtain an upper mass limit for neutron stars from this observation. The first phase of neutron production in red giants converts about 1/17 of its mass to mirror neutrons at maximum. For a $20 M_{\odot}$ star, therefore, it could end up with a core of $2.22 M_{\odot}$. If $2.22 M_{\odot}$ is indeed the limit, then stars need at least $20 M_{\odot}$ to collapse into black holes in the first phase. On the other hand, a star with $15\text{--}20 M_{\odot}$ can build a core up to $3\text{--}4 M_{\odot}$ during the second phase and then quietly turns into a black hole in the end. This may explains why the above-mentioned SNe II-L are so rare. Further studies on the mass limit of neutron stars and the nature of black holes can be found in Ref. \cite{tan2020b} based on supersymmetric mirror extensions of the new model \cite{tan2020,tan2020a}.

\section{Further implications of the new theory}

Now the interesting test mentioned in the introductory section \cite{mannarelli2018,goldman2019} can be easily answered. By the time the neutron star (more properly $n-n'$ star) forms, it is already evenly mixed between mirror and ordinary matter. So there is no mass loss or orbital period changing as suggested by Ref. \cite{mannarelli2018}. Therefore, the new theory is consistent with the test of pulsar timings and gravitational wave observations \cite{goldman2019}. The surprisingly low carbon content in rocky planets mentioned in Introduction could also be understood if these planets were formed from the ejected debris of type II-P supernovae.

Another interesting result that can be obtained under this theory is that oscillating movement from the mirror matter in the star is unavoidable as gravity serves as the restoring force for the oscillations. The oscillating period of the mirror matter can then be written as
\begin{equation}\label{eq_period}
\text{Period} = \sqrt{\frac{3\pi}{G\rho}}
\end{equation}
where $G$ is the gravitational constant and $\rho$ is the matter density where the mirror particles are located.
Due to the gravitational coupling, the ordinary matter has to do the counter movement and therefore presents some kind of pulsating behavior, in particular, periodic changes in luminosity. As a matter of fact, such behaviors are very common in stars, especially in red giants like the Cepheid variables that can be used to determine distances and the compact remnants like neutron stars \cite{vanderklis2006} and even white dwarfs \cite{nagel2004}. Such phenomena may help reveal the distribution and movement of mirror matter inside an astrophysical object or understand the laws for the mirror matter. For example, neutron stars have density of about $10^{14}$ g/cm$^3$ and an oscillation period of $\sim 10^{-3}$ s that can be estimated from Eq. (\ref{eq_period}) has indeed been observed in neutron stars \cite{vanderklis2006}. The 5-min oscillations from the Sun \cite{leighton1962,ulrich1970} could also be explained by a small amount of oscillating mirror matter in the center at a density of $\sim10^3$ g/cm$^3$. The typical period of a red giant variable is between hours and days that can be understood with oscillating mirror matter in its photosphere with a density of $1-10^{-3}$ g/cm$^3$ since, as discussed above, the variable star is constantly producing mirror neutrons that can migrate to the photosphere.

Such a pulsating behavior in the core that is evenly mixed with ordinary and mirror matter and new understanding of the core structures could shed light on the mechanism of supernova explosions \cite{janka2012a,burrows2013}. The large energy release of the neutron-conversion process from $n-n'$ oscillations near the core may also play a role. Meanwhile, the neutron-rich crust may provide an ample neutron source for the revived shock during a supernova explosion for synthesis of heavy elements via r-process. Taking into account new physics of this new star evolution theory, state-of-the-art supernova simulation models could potentially reveal how a core-collapse supernova is exploded.

\section{Conclusions}

To conclude, the new theory for single star evolution coupled with the $n-n'$ oscillation model is strongly supported by astrophysical observations. $^{13}$C($\alpha,n$)$^{16}$O and $^{17}$O($\alpha,n$)$^{20}$Ne are identified as the two critical nuclear reactions for the two-phase late stellar evolution as well as the free neutron sources for main and weak components of s-process, respectively. The mechanism of $n-n'$ oscillations plays an essential role in the formation of the stellar core with mirror matter.
Stellar nucleosynthesis, in particular, both s-process and r-process can be understood under the new theory. Progenitor sizes of compact stars and mass limits of neutron stars are also explained. Observed features of the two types of core-collapse supernovae match the predictions of the new mirror matter model well.
The mirror matter just like ordinary matter may indeed exist in our universe, especially in stars. 
This theory could also be applied to the studies for binary or multiple star systems. In particular, Type Ia supernovae, galaxy collisions \cite{markevitch2004,clowe2004}, and recently observed neutron star mergers \cite{ligoscientificcollaborationandvirgocollaboration2017} could be ideal for further test of this theory.

\begin{acknowledgments}
I would like to thank Ani Aprahamian and Michael Wiescher for supporting me in a great research environment at Notre Dame. I also thank Grant Mathews for pointing out the possibility of mirror neutrons escaping from the star.
This work is supported in part by the National Science Foundation under
grant No. PHY-1713857 and the Joint Institute for Nuclear Astrophysics (JINA-CEE, www.jinaweb.org), NSF-PFC under grant No. PHY-1430152.
\end{acknowledgments}

\bibliography{sn}

\end{document}